\documentclass[twocolumn]{aastex631}

\usepackage{makecell}

\usepackage{graphicx}
\usepackage{natbib}
\usepackage{amsmath}
\usepackage{mathtools}
\usepackage{subcaption}
\usepackage{url}

\usepackage{newtxtext,newtxmath}

\shorttitle{GRB polarimetry with {\em Astrosat}}
\shortauthors{Chattopadhyay et al.}

\begin{document}

\title{Hard X-ray polarization catalog for a 5-year sample of Gamma-Ray Bursts using {\em AstroSat} CZT-Imager}

\correspondingauthor{Tanmoy Chattopadhyay}
\email{tanmoyc@stanford.edu}

\author{Tanmoy Chattopadhyay}
\affiliation{Kavli Institute of Particle Astrophysics and Cosmology, Stanford University, 452 Lomita Mall, Stanford, CA 94305, USA}
\author{Soumya Gupta}
\affiliation{Inter-University Center for Astronomy and Astrophysics, Pune, Maharashtra-411007, India}
\affiliation{Homi Bhabha National Institute, Anushakti Nagar, Mumbai Maharashtra-400094, India}
\author{Shabnam Iyyani}
\affiliation{Inter-University Center for Astronomy and Astrophysics, Pune, Maharashtra-411007, India}
\affiliation{Indian Institute of Science Education and Research, Thiruvananthapuram, 695551, Kerala, India}
\author{Divita Saraogi}
\affiliation{Department of Physics, Indian Institute of Technology Bombay, Powai, Mumbai-400076, India}
\author{Vidushi Sharma}
\affiliation{Inter-University Center for Astronomy and Astrophysics, Pune, Maharashtra-411007, India}
\affiliation{Department of Physics, KTH Royal Institute of Technology, AlbaNova, 10691 Stockholm, Sweden}
\author{Anastasia Tsvetkova}
\affiliation{Ioffe Institute, Politekhnicheskaya 26, St. Petersburg 194021, Russia}
\author{Ajay Ratheesh}
\affiliation{INAF - IAPS, Via Fosso del Cavaliere 100, I-00133 Rome, Italy}
\author{Rahul Gupta}
\affiliation{Department of Physics, Deen Dayal Upadhyaya Gorakhpur University, Gorakhpur-273009, India}
\affiliation{Aryabhatta Research Institute of Observational Sciences (ARIES), Manora Peak, Nainital-263002, India}
\author{N.P.S. Mithun}
\affiliation{Physical Research Laboratory, Navrangpura, Ahmedabad, Gujarat-380009, India}
\author{C. S. Vaishnava}
\affiliation{Physical Research Laboratory, Navrangpura, Ahmedabad, Gujarat-380009, India}
\author{Vipul Prasad}
\affiliation{Inter-University Center for Astronomy and Astrophysics, Pune, Maharashtra-411007, India}
\author{E. Aarthy}
\affiliation{Physical Research Laboratory, Navrangpura, Ahmedabad, Gujarat-380009, India}
\author{Abhay Kumar}
\affiliation{Physical Research Laboratory, Navrangpura, Ahmedabad, Gujarat-380009, India}
\author{A. R. Rao}
\affiliation{Inter-University Center for Astronomy and Astrophysics, Pune, Maharashtra-411007, India}
\author{Santosh Vadawale}
\affiliation{Physical Research Laboratory, Navrangpura, Ahmedabad, Gujarat-380009, India}
\author{Varun Bhalerao}
\affiliation{Department of Physics, Indian Institute of Technology Bombay, Powai, Mumbai-400076, India}
\author{Dipankar Bhattacharya}
\affiliation{Inter-University Center for Astronomy and Astrophysics, Pune, Maharashtra-411007, India}
\author{Ajay Vibhute}
\affiliation{Inter-University Center for Astronomy and Astrophysics, Pune, Maharashtra-411007, India}
\author{Dmitry Frederiks}
\affiliation{Ioffe Institute, Politekhnicheskaya 26, St. Petersburg 194021, Russia}

\begin{abstract}
Cadmium Zinc Telluride Imager (CZTI) aboard AstroSat has been regularly detecting Gamma-Ray Bursts (GRBs) since its launch in 2015. Its sensitivity to polarization measurements at energies above 100 keV allows CZTI to attempt spectro-polarimetric studies of GRBs. Here, we present the first catalog of GRB polarization measurements made by CZTI during its first five years of operation. This presents the time integrated polarization measurements of the prompt emission of 20 GRBs in the energy range 100-600
keV. The sample includes the bright GRBs which were detected within an angle range of 0-60$^\circ$ and 120-180$^\circ$
where the instrument has useful polarization sensitivity
and is less prone to systematics. We implement a few new modifications in the analysis to enhance polarimetric sensitivity of the instrument. Majority of the GRBs in the sample are found to possess less / null polarization across the total bursts' duration in contrast to a small fraction of five GRBs exhibiting high polarization. The low polarization across the bursts can be speculated to be either due to the burst being intrinsically weakly polarized or due to varying polarization angle within the burst even
when it is highly polarized. In comparison to POLAR measurements, CZTI has detected a larger number of cases with high polarization. This may be a consequence of the higher energy window of CZTI observations which results in the sampling of smaller duration of burst emissions in contrast to POLAR, thereby, probing emissions of less temporal variations of polarization properties.
\end{abstract}

\keywords{X-rays: general --- Polarization --- gamma-ray burst: general --- instrumentation: detectors --- instrumentation: polarimeters}

\section{Introduction} \label{sec1}

GRBs \citep{discovery} are the sources of the brightest electromagnetic radiation known to occur in our Universe. These energetic events are believed to be powered by a newly born black hole \citep{woosley93, narayan2001} or a magnetar \citep{Usov1992, Duncan_Thompson1992}, formed during the core-collapse of a massive star \citep{woosley93, iwamoto98, macfadyen99} or the merger of compact objects \citep{eichler89, narayan92} such as binary neutron stars \citep{Abbott_170817A} or black hole and neutron star. The initial intense flashes of $\gamma$-rays occurring close to the burst site is known as prompt emission. The delayed emission which is observed across the entire energy spectrum is known as afterglow \citep{Rees1992, piran05, meszaros06}.

The unique and non-recurring transient nature of the GRB emission makes the task of developing a generic understanding of the GRB radiation mechanism highly challenging. Even after more than half a century since its discovery, the mechanism giving rise to the observed prompt emission still largely remains an open key question \citep{kumar15, zhang19}. Generally, the spectrum of the prompt emission is studied via fitting with various phenomenological model functions like power law, Band function, power law with exponential cutoff etc \citep{band93,gruber2014fermi}. The most popular competing radiation models are based on the jet photosphere \citep{Rees1992, ryde04} and optically thin synchrotron emission \citep{rees94,sari98}. Several attempts have been made to test these various physical models directly with data \citep{Ahlgren_etal_2015,Ahlgren_etal_2019,Burgess2014,Burgess_etal_2020}.  
However, most spectral analyses  
yield similar fit statistics for different empirical and physical models \citep{iyyani15, Zhang_2016}. Thus, spectral analysis alone leads to ambiguity in selecting the best fit model. One of the ways forward is to have more constraining observables like polarization which can help to break 
this degeneracy \citep{toma08, covino16, mcconnell16, gill18, Gill_etal_2021_review}. This is possible because different radiation models predict different ranges of polarization. For instance, one expects low to null polarization for photospheric emission while a wider range of polarization values is expected for synchrotron radiation depending on the viewing geometry and magnetic field configurations present at the emission site \citep{waxman03,granot03,toma08}.
Therefore, polarization measurements along with spectrum could reveal a clearer picture of the underlying radiation mechanism. In addition, statistical studies of a large sample of GRB polarization values would help assess the possible generic radiation model in GRBs.

In the last two decades, several hard X-ray spectrographs as well as dedicated X-ray polarimeters 
have contributed to polarization measurements of GRB prompt phase. A summary of the GRB polarization measurements can be found in \citet{mcconnell16}, \citet{chattopadhyay21_review} and \citet{Gill_etal_2021_review}. However, statistically significant polarization measurements are available only for a handful of GRBs and the obtained values exhibit a wide range. 
For example, GAmma-ray Polarimeter \citep[GAP,][]{gap2011} onboard {\it IKAROS} provided polarization measurements for three bright GRBs
\citep{yonetoku11,yonetoku12}. A high polarization fraction was reported for GRB 110301A ($\sim$70\%) and GRB 110721A ($\sim$84\%) in the full burst period, whereas, in case of GRB 100826A, a flip in polarization angle was seen between two pulses.   
Recently, POLAR \citep{produit18} provided precise polarization measurements for 14 GRBs \citep{zhang19,Kole20polar_catalog}. They, however, reported low level of polarization across the burst duration for majority of the GRBs in their sample.
For GRB 170114A, a temporal evolution of polarization was seen within the pulse \citep{zhang19,Burgess_etal_2019}.
CZTI onboard {\em AstroSat} reported high polarization for most of the 11 bright GRBs detected in the first year of its operation \citep{chattopadhyay19}. 
The detailed spectro-polarimetric analysis of some of these bright GRBs revealed time and energy-dependent variation in polarization happening across the burst duration \citep{chand18a,chand18b, Sharma_etal_2019, Sharma20}. Most importantly, we note that the detailed study of GRB 160821A showed a $\sim$90$^\circ$ flip in polarization angle happening twice within a single broad emission pulse \citep{Sharma_etal_2019}. Thus, the temporal evolution of polarization observed within the bursts, GRB 160821A and GRB 170114A, suggested that the prompt emission was highly dynamic resulting in the dilution of the observed polarization measured across the total burst duration. Therefore, the existing polarization measurements indicated that the polarization in GRBs spans a wide range from low to high values. However, a preference towards low polarization values was observed in the sample and it would be interesting to classify them into cases of intrinsic low polarization or concealed change in angle.

While POLAR has stopped its scientific operations after 2017, CZTI on board {\em AstroSat} continues to work as a sensitive GRB monitor. Here, we present the polarization catalog for $20$ bright GRBs that were observed in the first five  years  of the operation of CZTI (2015 October 6 
to 2020 October 5).
In this sample, besides selecting the GRBs with fluence higher than 10$^{-5}$ erg cm$^{-2}$, we choose only those GRBs that were detected within the incident angle range of $\sim$0-60$^\circ$ and $\sim$120-180$^\circ$ in CZTI co-ordinates. The sample selection and the observed properties of the GRB sample are discussed in Sec. \ref{sample}. We have also
improved the GRB polarization analysis technique by including better noise rejection algorithms and by enhancing the polarimetric sensitivity at higher energies. The details of the data analysis and the new improvements have been discussed in Sec. \ref{method}. 
In Sec. \ref{results}, we discuss the polarization results for the sample followed by discussions in Sec. \ref{discussion}  and conclusions in Sec. \ref{Conclusions}. 

\section{The 5 year sample of GRBs for polarimetry }\label{sample}
CZTI has been working as a prolific GRB monitor since the launch of {\em AstroSat}. In the  five years selected for the present study, between 2015 October 6 and 2020 October 5, CZTI detected a total of 413 GRBs\footnote{\url{http://astrosat.iucaa.in/czti/?q=grb}}. 
For polarization analysis, we selected 126 GRBs with fluence above $10^{-5}$~erg~cm$^{-2}$ in 10-1000 keV.
The {\em Fermi} and {\em Swift} catalogues list 279 GRBs (41 GRBs are in common) above this fluence limit during the corresponding 5 year period. Considering $\sim 30$\% of Earth occultation at any given time, and $\sim 18$\% of data gap due to passage through the South Atlantic Anomaly (SAA) or telemetry errors, detection of the  bright GRBs in CZTI ($\sim$ 44\%) is mostly consistent with its effective duty cycle and sky coverage.
We, however,  found that some fraction of far off-axis GRBs ($>$60$^\circ$) is lost due to obscuration  by other payloads, e.g., LAXPC, SXT, UVIT, and the spacecraft structures around CZTI. However, the detection is close to 100\% for  GRBs detected in 0-60$^\circ$ implying that the sample is complete in this angle range.  

A subset of 77 GRBs in the list of 126 GRBs are detected in $\sim$0-60$^\circ$ and $\sim$120-180$^\circ$ angle range in CZTI co-ordinates which is required for polarization analysis (more details in subsec. \ref{grbs}). 
We also impose a minimum polarimetric sensitivity (MDP or Minimum detectable Polarization) of 40\% to the GRBs, based on the detected number of Compton scattered photons,  which limits the final sample to 20 GRBs for polarization study. 

\subsection{Selection of GRBs}\label{grbs}

Pixelated CZT detectors having the ability to record ionizing events in neighbouring pixels simultaneously can be used to identify Compton scattered events and hence are capable of measuring polarization properties of incoming radiation in the hard X-ray regime (above 100 keV). CZTI has been calibrated with polarized on-axis sources before the launch of {\em AstroSat} \citep{chattopadhyay14,vadawale15}, thus demonstrating that subtle systematics involving reading data from neighbouring simultaneous events are well understood and taken care of. For off-axis sources like GRBs, it was assumed that we could use the physics of radiation interaction to extract the polarization information \citep{chattopadhyay19}. However, additional systematics could be present while using planar detectors to measure the scattering angles from off-axis sources. To particularly address these issues, we conducted an experimental study with a spare CZT detector module (identical to the one used in {\em AstroSat}-CZTI) to investigate the
polarization sensitivity of CZTI for off-axis polarized radiation (Vaishnava et al., under review).
The experiment was conducted with a partially polarized 190-240 keV continuum at various off-axis polar and azimuthal angles with respect to the CZT module to compare the measured polarized fraction with that of incident X-rays.
Because of the complicated geometry and diverging X-rays, it was difficult to estimate the polarization fraction of the incident X-rays analytically.
Hence we carried out Geant4 \citep[GEometry ANd Tracking,][]{agostinelli03} simulations of the complete experimental setup starting from the generation of a partially polarized source from unpolarized X-rays from the $^{133}$Ba source and estimated the polarization fraction of the X-rays incident on the detector by recording the electric field vector for each
photon. We then analyzed the experimental and simulated data using the same techniques. We found that the measured polarization fraction from experiments and simulations agree with each other and with the incident polarization fraction within statistical uncertainties. For larger incidence angles (45-60$^\circ$), however, it was found that the assumption of approximating a divergent beam with a parallel beam could induce some experimental systematics,
giving a slight mismatch between the estimated and original polarization (this should not be present in cosmic sources like GRBs).  
At angles beyond $\sim$60$^{\circ}$ from the detector normal, polarization sensitivity is shown to be relatively low (i.e. MDP is high) compared to that for lower inclinations.
These results helped fine-tune the analysis methods of polarization, particularly the selection of GRBs for polarization analysis.

We used the $\mu_{100}$ values (modulation amplitude for 100\% polarized radiation) obtained from the Geant4 simulation of the {\em AstroSat} mass model to calculate MDP values for the GRBs from the known GRB and background count rates. A $\sim$40\% limit on the MDP level was imposed to select the GRBs for polarization analysis. This criteria in most of the cases flags out the GRBs detected at off-axis angles between  60$^\circ$ and  120$^\circ$ and leaves 20 GRBs out of 77. 

The list of the twenty GRBs and their observed properties are given in Table \ref{table1}. In the first column under the name of the GRBs, we specify the instruments that were triggered. Fourteen of the GRBs have triggered {\em Fermi}/GBM detectors, seven of them have triggered {\em Swift}/BAT detectors while two of them triggered both these detectors. 
There are four GRBs triggered in both {\em Swift}/BAT and {\em Konus}-Wind. There are two GRBs, GRB 200806A and GRB 190928A, which triggered only BAT and {\em Konus}-Wind respectively. 
Localization information for GRB 190928A is available from IPN. The GRB location error circles quoted in the table are taken from {\em Swift}/XRT, {\em Swift}/BAT and {\em Fermi}/GBM catalogs or the respective {\em GCNs}. 
The last column of 
Table \ref{table1} shows the respective polar 
($\theta$) and the azimuthal ($\phi$) angles of detection in CZTI coordinates.
It is to be noted that in \citet{chattopadhyay19}, we reported polarization measurements for eleven bright GRBs from the first year of {\em AstroSat}. However, in this work, polarization analysis has been conducted only for six of those GRBs as the rest were detected at far off-axis angles and do not satisfy the 40\% MDP criteria. 

\subsection{Selection of burst interval}\label{time}
In this work, polarization analysis has been conducted for the time integrated emission of the bursts. 
The burst interval is chosen by employing Bayesian block algorithm \citep{scargle1998studies,Scargle2013,Burgess2014} on the 2-pixel Compton event light curves of the bursts (definition of the Compton events and their selection are discussed in the next section). 
In Bayesian block analysis, the probability density corresponding to the background region is used to identify the start and stop times of the burst. The time stamp of the first block close to the burst with probability density greater than that of the background is considered as the start time of the GRB. The stop time is estimated in a similar way. 
The start and stop times ($t_1$ and $t_2$) for all the bursts and the burst duration used for polarization analysis are given in Table \ref{table1}. For the GRBs detected by {\em Fermi}, $t_1$ and $t_2$ are given with respect to the GBM trigger times. For the remaining GRBs, the trigger times were taken either from {\em Swift}-BAT or {\em Konus}-Wind. 

\begin{table*}%
\centering
\begin{tiny}
\caption{The sample of GRBs selected for polarization study with CZTI }


\begin{tabular}{p{1.1cm}ccp{1.45cm}cccp{1.3 cm}cp{1.2 cm}p{1.2cm}}
\\\hline\\[0.03cm]

GRB  &Localization${}^{a}$& Burst Interval (s)${}^{b}$& ($t_1$, $t_2$)${}^{c}$ & $\alpha$ &$\beta$   & $E_p$/ $E_c$  &  Fluence${}^{b}$ ${}^{d}$ (10-1000 keV) & Incident Direction ($\theta, \phi$) \\

(Detectors) &   & (s)  &     (s)    &    &   &  (keV)  & ($\times10^{-5}$ erg/cm$^{2}$) &($^\circ$)  \\[0.03cm]\\ \hline\\
160325A (GBM, BAT) &$1^{\arcsec}.7$& 43.82 (42.94) & (2.28,46.10) & $-0.75_{-0.070}^{+0.084}$& $-1.97_{-0.140}^{+0.100}$& $223.57_{-25}^{+29}$ & 2.00 (1.86) & 0.66,159.48 \\[0.03cm]

160623A$^{*}$ (GBM, Konus-Wind)& $3^{\arcsec}.5$ & 17.05 (107.78) & (1.16,18.21) & $-0.94_{-0.020}^{+0.018}$& $-2.83_{-0.100}^{+0.090}$ & $662_{-18}^{+19}$ & 39.3 (0.39) & 140.52,118.09\\[0.03cm]

160703A$^{*}$ (BAT, Konus-Wind)&$3^{\arcsec}.9$ & 24.91 (44.40) & ($-$1.78,23.12) & $-0.78_{-0.090}^{+0.120}$& $<-2.48$& $351_{-46}^{+40}$& 2.02 (0.90) & 10.15,95.08 \\[0.03cm]

160802A (GBM)&$1^{\circ}.0$ & 18.07 (16.38)& (0.03,18.11)  &$-0.64_{-0.030}^{+0.040}$& $-2.53_{-0.200}^{-0.140}$& $207_{-1}^{+1}$& 6.36 (6.84) & 52.95,273.12 \\[0.03cm]

160821A (GBM, BAT) &$1^{\arcmin}.0$& 37.96 (43.01)& (117.18,155.135)& $-0.96_{-0.003}^{+0.003}$ & $-2.29_{-0.015}^{+0.015}$ & $977_{-12}^{+12}$ & 48.1 (52.22) & 156.18,59.27 \\[0.03cm]

170527A (GBM)& $1^{\circ}.0$ & 37.95 (49.15)& ($-$0.76,37.18) &$-0.99_{-0.011}^{+0.011}$& $-3.14_{-0.290}^{+0.290}$& $974_{-47}^{+51}$& 8.36 (8.43) & 26.54,101.57 \\ [0.03cm]

171010A (GBM)&$1^{\arcsec}.4$ & 99.06 (107.27) & (7.14,106.20) & $-1.12_{-0.002}^{+0.005}$& $-2.39_{-0.024}^{+0.024}$& $180_{-3}^{+3}$& 63.1 (63.28) & 55.30,35.19 \\[0.03cm]


171227A (GBM)&$1^{\circ}.0$ & 30.01 (37.63) & (0.26,30.27)  &$-0.80_{-0.011}^{+0.011}$& $-2.49_{-0.049}^{+0.047}$& $899_{-32}^{+32}$& 26.8 (28.96)  & 146.49,353.57 \\[0.03cm]

180103A$^{*}$ (BAT, Konus-Wind) &$3^{\arcsec}.8$& 165.83  & (11.38,177.21) & $-1.31_{-0.060}^{+0.060}$& $-2.24_{-0.130}^{+0.900}$& $273_{-23}^{+26}$& 22.3  & 52.33,108.17\\ [0.03cm]

180120A (GBM)&$1^{\circ}.0$ & 24.01 (28.93) & (0.09,24.10) & $-1.01_{-0.014}^{+0.014}$& $-2.40_{-0.090}^{+0.090}$& $140.91_{-3}^{+3}$& 5.68 (6.45) & 15.89,206.28 \\[0.03cm]

180427A (GBM)&$1^{\circ}.0$ & 13.01 (25.92)& (0.15,13.16)  &$-0.29_{-0.077}^{+0.077}$& $-2.80_{-0.160}^{+0.160}$& $147_{-2}^{+2}$& 4.34 (5.04) & 40.81,257.79 \\[0.03cm]

180806A (GBM) &$5^{\arcsec}.0$& 10.33 (15.62)& ($-$0.01,10.32) & $-0.92_{-0.036}^{+0.039}$& $-2.46_{-0.440}^{+0.230}$& $453_{-44}^{+46}$& 1.78 (2.24) & 26.79,246.11\\[0.03cm]

180809B$^{*}$ (BAT, Konus-Wind)&$1^{\arcsec}.4$ & 63.00 (233.20) & ($-$3.89,59.11) & $-0.69_{-0.070}^{+0.080}$& $-2.29_{-0.080}^{+0.070}$& $251_{-15}^{+16}$& 23.6 (7.30) & 61.44,17.49 \\[0.03cm]

180914A (GBM)&$5^{\arcsec}.0$ & 128.01 (122.37) & (5.34,133.35)  &$-0.73_{-0.031}^{+0.032}$& $-2.30_{-0.150}^{+0.110}$& $330_{-19}^{+20}$& 9.76 (8.19) & 52.92,173.40 \\[0.03cm]

180914B$^{*}$ (BAT, Konus-Wind) &$3^{\arcsec}.4$& 170.04 & ($-$13.81,156.23)& $-0.75_{-0.040}^{+0.040}$ & $-2.10_{-0.700}^{+0.080}$& $453_{-24}^{+26}$ & 59.8  & 36.45,314.92\\ [0.03cm]


190530A (GBM)& $1^{\arcsec}.4$ & 26.86 (18.43) & (7.24,34.10) & $-0.99_{-0.002}^{+0.022}$& $-3.50_{-0.250}^{+0.250}$& $888_{-8}^{+8}$& 38.5 (37.06)  & 154.50,79.87 \\[0.03cm]

190928A$^{*}$ (Konus-Wind)&$2^{\circ}.1$ & 119.97 & ($-$2.70,117.26)  &$-1.00_{-0.060}^{+0.060}$& $-1.97_{-0.130}^{+0.070}$& $658_{-88}^{+111}$& 20.5 & 57.69,231.26\\[0.03cm]


200311A (GBM) &$1^{\circ}.0$& 39.07 (52.48) & (0.14,39.20)& $-0.95_{-0.020}^{+0.020}$& $-2.57_{-0.190}^{+0.190}$& $1218_{-110}^{+110}$& 114 (4.25) & 29.89,151.24\\[0.03cm]

200412A (GBM) &$1^{\circ}.4$ & 14.17 (12.61) & ($-$0.89,13.28) & $-0.70_{-0.050}^{+0.050}$& $-2.50_{-0.210}^{+0.210}$& $256_{-7}^{+8}$ & 3.24 (2.87) & 41.57,272.66 \\[0.03cm]

200806A (BAT) & $1^{\arcsec}.7$ & 38.04 (38.80) & (-2.85,35.18) & -0.53 & -2.96 & 109.12 & 2.44 (0.10) &  6.78,262.58\\[0.03cm]\\
\hline
\end{tabular}

$^a$Localization given with 90\% error radius, taken from Swift/XRT, Swift/BAT, and {\em Fermi}/GBM catalogs. For GRB 190928A, localization information is available from
IPN\\

$^*$Spectral parameters obtained from {\em Konus}-Wind for the following burst interval 1. GRB160323A: 18.176 s, 2. GRB160703A: 24.832 s, 3. GRB180103A: 169.984 s, 4. GRB180809B: 64.256 s, 5. GRB180914B: 160.000 s and 6. GRB190928A: 119.808 s

$^b$ The values inside the parenthesis are the T90 and fluence value in the Burst Interval and fluence column respectively as reported in the IceCube catalogue \url{https://user-web.icecube.wisc.edu/~grbweb_public}.

$^c$t$_1$ and t$_2$ are w.r.t. GBM/BAT/{\em Konus}-Wind  trigger-time\\

$^{d}$Fluence is in the range $t_1$ to $t_2$ in 10-1000 keV. \\

\label{table1}
\end{tiny}
\end{table*}

\subsection{Spectroscopic properties of the sample}\label{spectra}
Knowledge of spectral energy distribution of the incoming photons is critical to model the interaction of the photons with the spacecraft, instrument supporting structures and other payloads. Geant4 platform is normally used to model these effects, given that the energy distribution of photons for a given GRB is known.  
To extract the spectral parameters for all the selected GRBs at the same time intervals as have been used for the polarization measurements, we carried out broadband spectroscopy using all the available data. 

For the thirteen GRBs (see Table \ref{table1}) detected by GBM (we do not analyze GRB 160623A with GBM as the prompt emission was detected only partially at a later phase of the burst), the GBM time tagged events were retrieved from the {\em Fermi} Science Support Center
archives\footnote{\url{https://fermi.gsfc.nasa.gov/ssc/data/access/}}. 
The photon spectra were fitted with  Band model \citep{band93}.
For the six GRBs detected by {\em Konus}-Wind, we fitted {\em Konus}-Wind data with Band function to estimate the spectral parameters \citep{Svinkin_2016,anastesia17,anastesia21}. 
Since the typical accumulation time for a KW ({\em Konus}-Wind) spectrum is around 8 s, it is not possible to accumulate the data precisely in the selected time intervals. For spectral analysis, the closest time stamps to $t_1$ and $t_2$ were used (see footnote of Table \ref{table1}). 
For GRB 200806A detected by BAT, we obtained the spectral parameters from BAT and CZTI data analysis. The background-subtracted BAT spectral files in the energy range $15-150\, \rm keV$ were extracted from the event file using the standard procedure\footnote{\url{https://swift.gsfc.nasa.gov/analysis/threads/bat_threads.html}}. Because of the limited energy window of the BAT spectral data, we used the CZTI spectral data that includes the single pixel (100-900 keV), 2-pixel Compton events (100-700 keV) and CsI-Veto events (100-500 keV) for a combined spectral analysis. The CZTI spectral files were generated and analysed using the methodology described in \cite{chattopadhyay21_grb}.  
The obtained spectral parameters ($\alpha$, $\beta$) and peak energy ($E_p$) are tabulated in Table \ref{table1}. The quoted errors on the parameters are for 90\% confidence level. From the spectral parameters, we calculated the fluence in the selected time intervals in 10--1000 keV band (given in the eighth column of the table). Values of the derived  spectral parameters are consistent with those reported in the GCNs and catalogs for the full burst interval. 

\section{Methodology of GRB polarimetry with CZTI} \label{method}
CZTI consists of an array of 64 CZT detector modules where each detector is 
5 mm thick and spatially segmented into 256 pixels (with a nominal 
pixel size of 2.5 mm $\times$ 2.5 mm). Availability of photon tagging mode in CZTI (time resolution of 20 $\mu$s) 
and significant 
Compton scattering cross-section beyond 100 keV 
enables CZTI to work as a sensitive Compton polarimeter.
\citet{vadawale17} reported the measurement of polarization of Crab pulsar and nebula using CZTI in 100-380 keV energy band. Later, using the same principle, polarization measurement for a sample of GRBs was reported by \citet{chattopadhyay19}, where the details of the method are also presented. 
Here we briefly summarize the steps. 
\begin{itemize}
    \item We identify the adjacent 2-pixel events from 20 $\mu$s coincidence window.
    \item We demand that the ratio of the energies deposited in two adjacent pixels is between 1 and 6 to filter out the background and noisy events. Geant4 simulation study shows that the signal to noise is optimum in this ratio range.
    \item These steps are applied on both the burst region obtained from the Bayesian block analysis on the GRB lightcurves (see Sec. \ref{time}) and at least 300 seconds of pre-burst and post-burst background intervals. For each valid event, we estimate the azimuthal scattering angle. The azimuthal angle distribution from the list of valid events in the background region is subtracted from that of the GRB region.
    \item An unpolarized azimuthal angle distribution is then used to correct the background subtracted azimuthal distribution for the inherent modulation that is seen even for fully unpolarized radiation \citep{chattopadhyay14}. The square pixels in CZTI detector plane introduce an asymmetry in the scattering geometry which causes the observed inherent modulation. The unpolarized distribution is obtained from the {\em AstroSat} mass model \citep{mate21} by simulating 10$^9$ unpolarized photons in Geant4 with source photon energy distribution same as the GRB spectral distribution (modeled as Band function) and for the same orientation with respect to the spacecraft. 
\end{itemize}
The corrected azimuthal angle distributions are then fitted with a sinusoidal function to calculate the modulation amplitude ($\mu$) and polarization angle in the detector plane ($\phi_0$) using Markov Chain Monte Carlo  (MCMC) method. 
There is an alternative method (`template fitting method') where the raw histograms can be fitted with templates generated from simulations. The modulation curve fitting method assumes cosine nature of the azimuthal distributions, which is not strictly true at larger angles, particularly when the distributions have
high counting statistics.
The off-axis experiment with CZTI detectors (Vaishnava et al., 2022, under review) shows that modulation curve fitting provides same results as template fitting even for Compton events, a factor of ten larger than typical number of events found in case of GRBs. Modulation curves help visual representation and since the same method has been used in earlier analysis, for consistency, we continue with the standard sinusoidal method in this work.

Polarization fraction is obtained by normalizing the fitted $\mu$ with $\mu_{100}$, where $\mu_{100}$ values are obtained by simulating the {\em Astrosat} mass model in Geant4 for 100\% polarized radiation (10$^9$ photons) for the same GRB spectral distribution and orientation. For each GRB, we calculate Bayes factor for confirmation of detection of polarization. For GRBs with Bayes factor $<$~3, we estimate upper limits in polarization. Details of the Bayes factor and upper limit calculations are given in \citet{chattopadhyay19}. 
 
\subsection{New improvements in the analysis}\label{improvements}
There are some new developments in the CZTI polarimetry analysis, particularly in case of GRBs, that improve the overall polarimetric sensitivity of CZTI and help in taking care of the systematic effects. Here we discuss these new developments briefly.

\subsubsection{Extending the polarimetry energy range to 600 keV}
Recently, in \citet{chattopadhyay21_grb}, we discussed calibration of a fraction of CZTI pixels (around 20\% of the CZTI pixels) that were found to have lower gains from on board calibration soon after the launch of {\em AstroSat} (we will refer to these pixels as low gain pixels hereafter).
Because of lower gains, these pixels have a higher energy
threshold of $\sim$60 keV for X-ray photon detection but are also sensitive to photons of much higher energies up to $\sim$1 MeV.
We used  five years of CZTI background data to calibrate these low gain pixels using 
high energy particle induced Tellurium activation lines at 88 keV and 145 keV \citep{odaka18} and by comparing the count distribution between the low gain and the normal gain CZTI pixels.       
The low gain pixels were thereafter successfully utilized in the CZTI analysis software to obtain single pixel spectra and 2-pixel Compton spectra in $\sim$100-1000 keV for a sample of eleven bright GRBs \citep[for details, see][]{chattopadhyay21_grb}. 
We also did not see any variation in gain of these pixels over time.
Here we explore the possibility of using the low gain pixels in extracting polarization information. 
Use of the low gain pixels enhances the signal to noise ratio of the signal because of the availability of additional collecting area. However, these additional events (combination of two low gain pixels or one low gain and one normal gain pixel) are mostly associated with high energy photons and at those energies, the distinction of the first (scattering of the incident photon) and second event (subsequent absorption of the scattered photon) can be sometimes difficult because for incident photon energies $>$260 keV, energy deposited in the first event can be larger than that in the second event for some of the scattering angles. 

We attempt to address this problem using Compton kinematics and Monte Carlo simulations. 
From the known Compton kinematics formulation, for each observed Compton event, we first calculate $\theta_{critical}$ for $E_{total}~ (=E_{low} + E_{high}) > 260$ ~keV, such that the scattered photon energy or the energy deposited in the second event is lower than the electron recoil energy or the energy deposited in the first event. 
This is followed by the calculation of the polar angle of scattering ($\theta$) from $E_{total}$ assuming that $E_{low}$ is the scattered photon energy or the energy deposited in the second event. If $180^\circ~>~\theta~>~\theta_{critical}$, we calculate the probability of such scattering scenario by integrating the Compton scattering cross-section from $\theta$ to 180$^\circ$. If the probability is high ($Probability~>R[0,1]$, a random number drawn from a uniform distribution between 0 and 1), the low energy deposition is classified as the second event and used to calculate the azimuthal angle of scattering correctly.   
Although the low gain pixels are sensitive to very high energies, here we impose a 600 keV limit because polarization analyzing power (represented by $\mu_{100}$) of CZTI or CZTI like Compton geometry decreases significantly at higher energies.

\begin{figure}
\centering
\includegraphics[scale=.5]{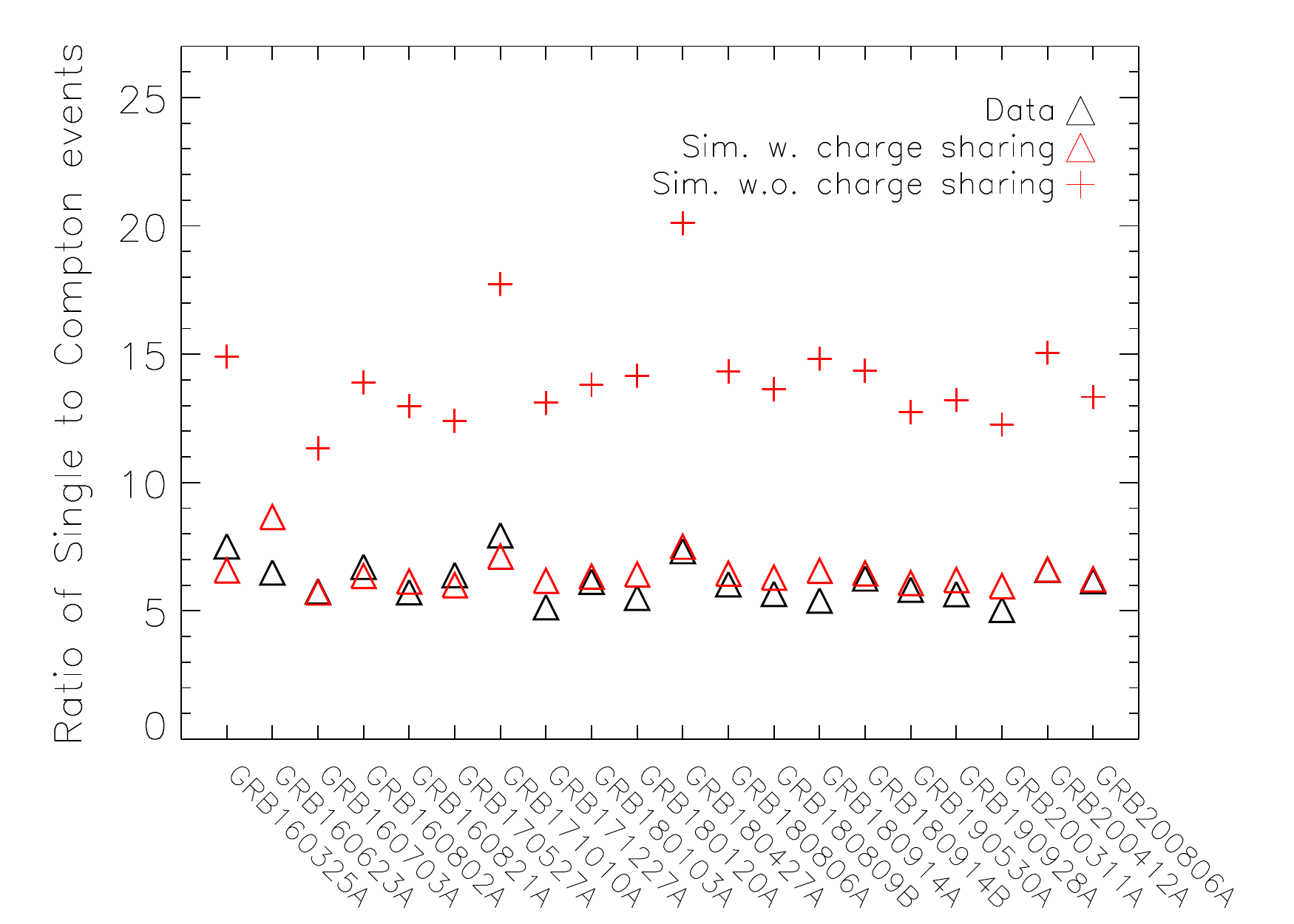}
\caption{Comparison of ratio of single to Compton events from observed data and simulation. The observed data points are shown in black triangles. If we do not include charge sharing in simulation, simulated ratios (red cross) are a factor 2 higher. However, after charge sharing correction, the simulated values shown in red triangles agree closely with the observed values.}
\label{fig_events}
\end{figure} 

\subsubsection{Improvement in noise rejection}
There is an ongoing effort to improve upon the existing noise rejection algorithms for CZTI. \citet{ratheesh21} summarize the new improvements in the background and noise rejection methods. One of the important developments, particularly for the polarization analysis, is the implementation of `Compton noise' algorithm. Some of the adjacent noisy pixels flickering at time scale less than 20 $\mu$s are seen to filter into the Compton event list and cause systematic artifacts in the azimuthal angle distribution. In the new noise rejection algorithm, we identify these events in the 2-pixel Compton detector plane histogram as outliers and filter them out with a user defined outlier level (n $\sigma$). This is carried out separately for the edge and the corner pixels.
Further details on the Compton noise analysis can be found in \citet{ratheesh21}.
We generate the Compton event list for all the twenty GRBs after applying the 2-pixel noise correction. 

\subsubsection{Charge sharing effect in CZTI}
 
\begin{figure*}
\centering
 \begin{subfigure}[b]{0.4\textwidth}
\includegraphics[width=\textwidth]{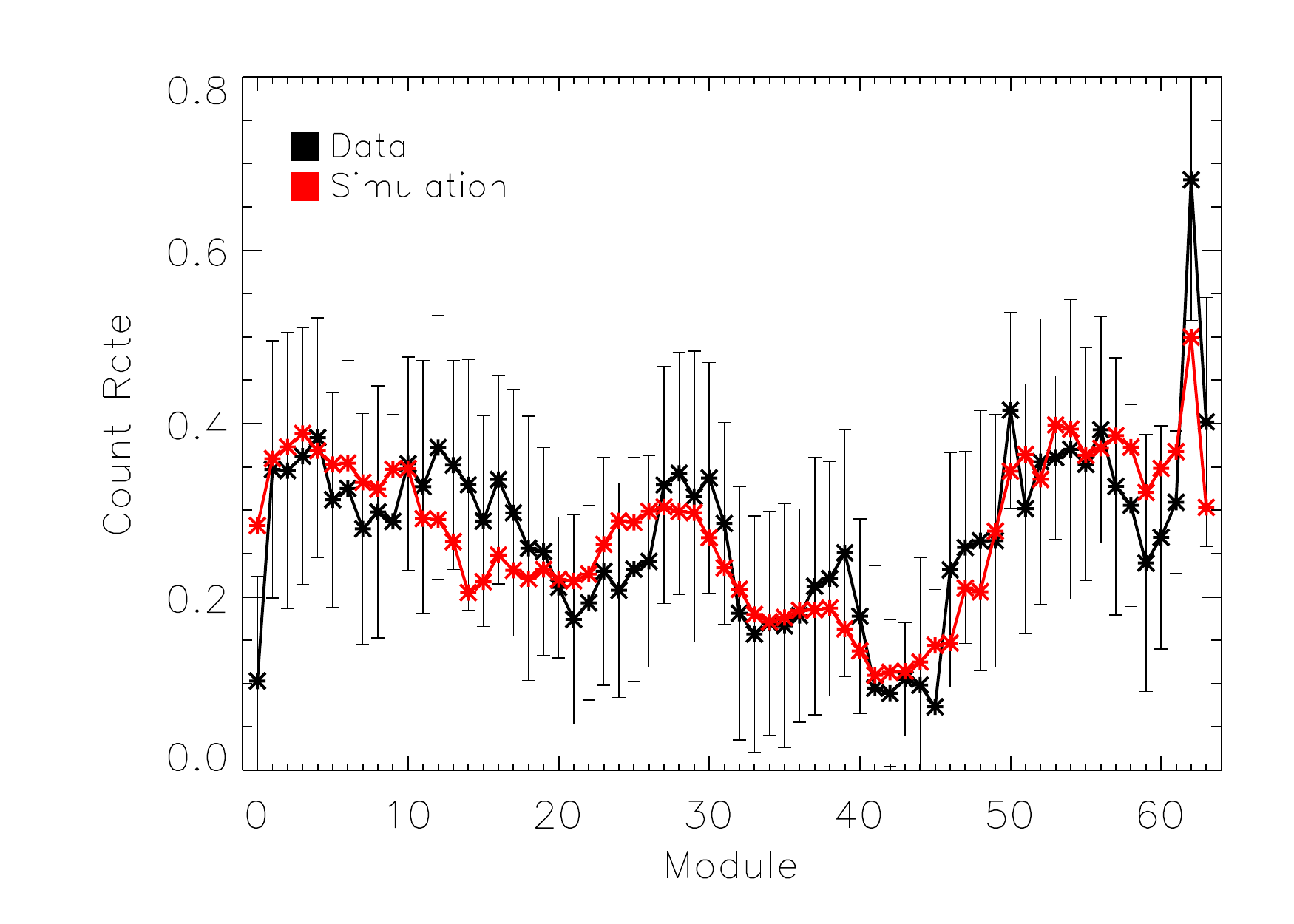}
\caption{GRB 160325A: $\theta = 0.66^\circ$, $\phi = 159.48^\circ$}
\end{subfigure}
\begin{subfigure}[b]{0.4\textwidth}
\includegraphics[width=\textwidth]{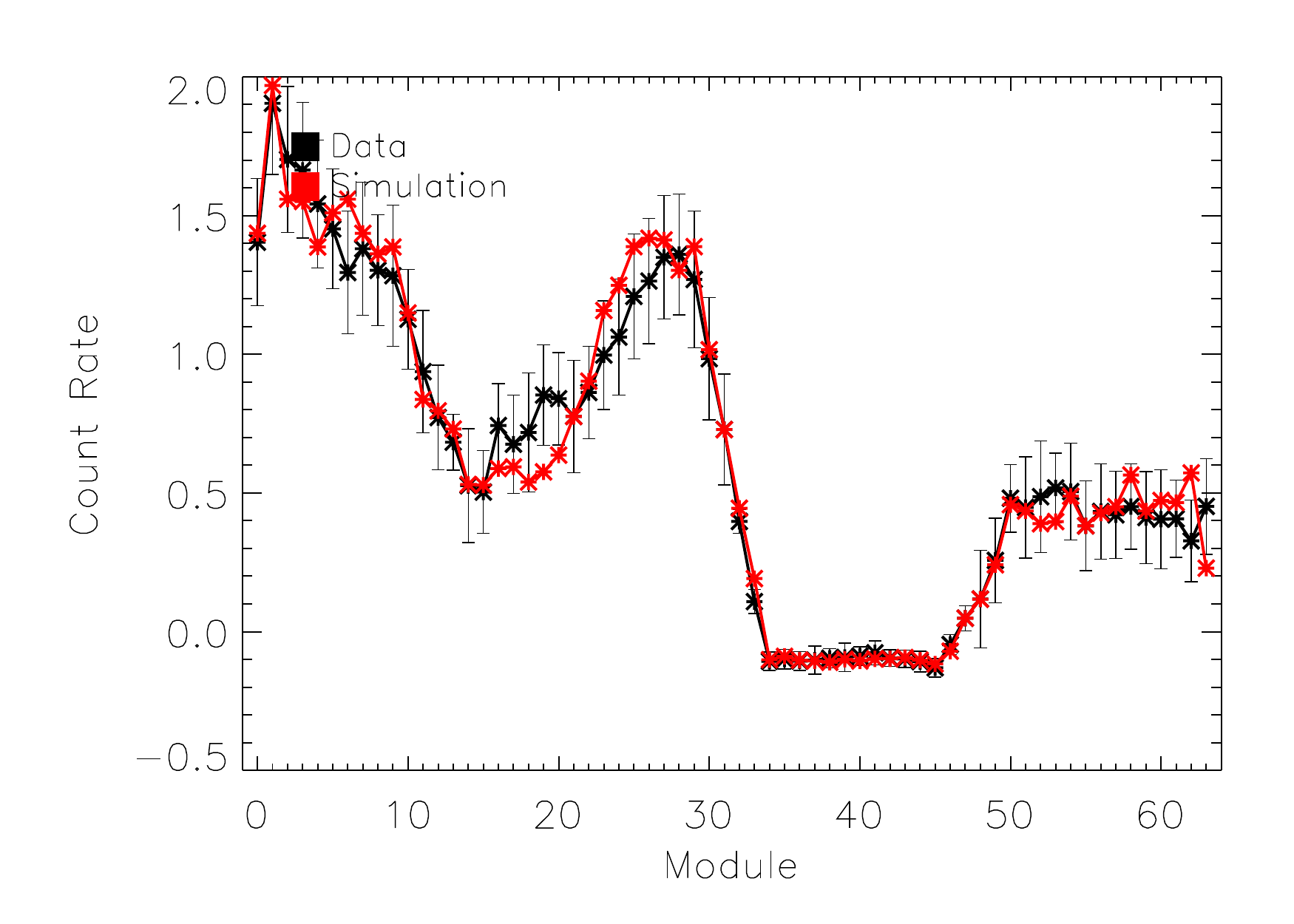}
\caption{GRB 170527A: $\theta = 26.54^\circ$, $\phi = 101.57^\circ$}
\end{subfigure}
\begin{subfigure}[b]{0.4\textwidth}
\includegraphics[width=\textwidth]{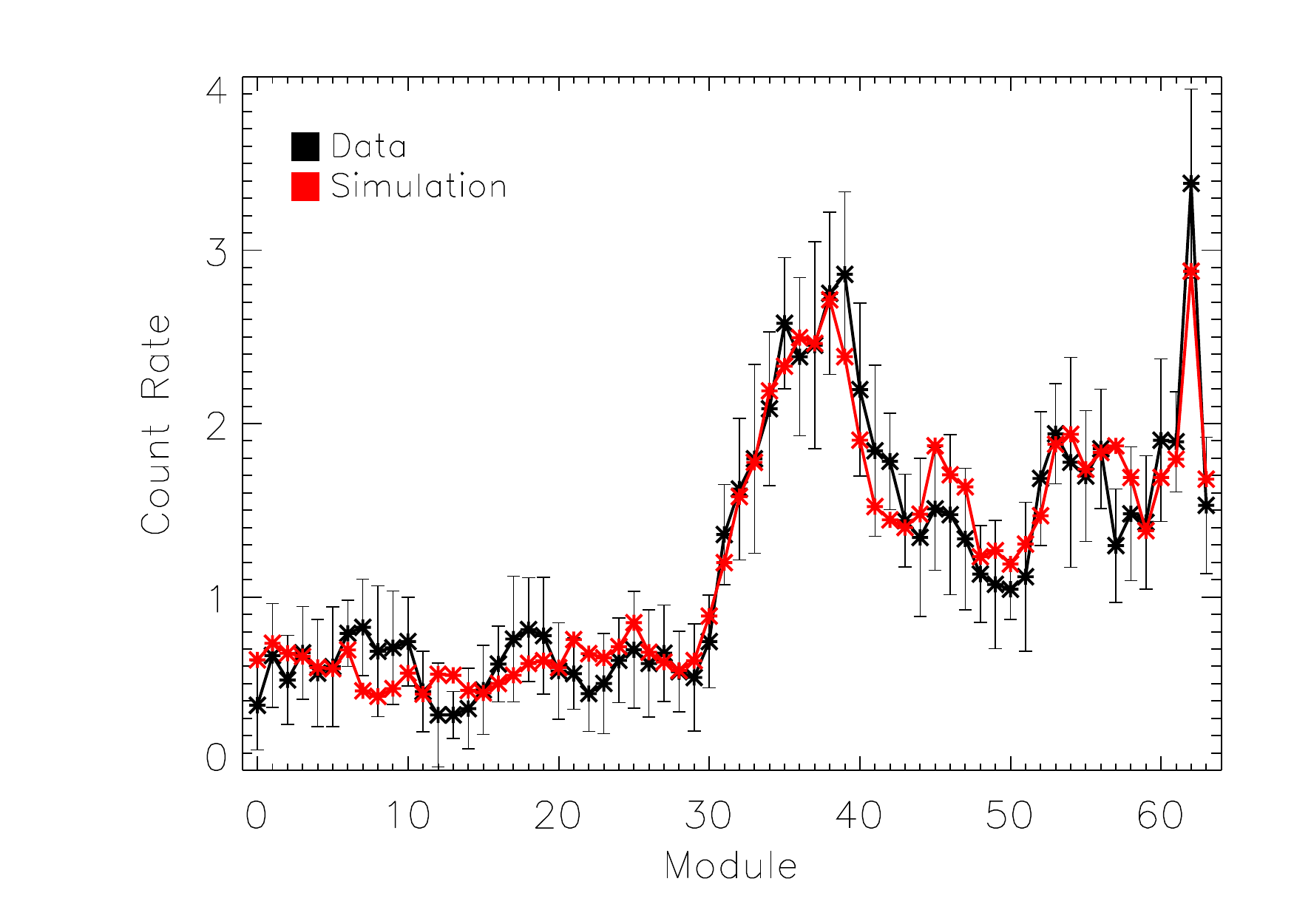}
\caption{GRB 180427A: $\theta = 40.81^\circ$, $\phi = 257.79^\circ$}
\end{subfigure}
\begin{subfigure}[b]{0.4\textwidth}
\includegraphics[width=\textwidth]{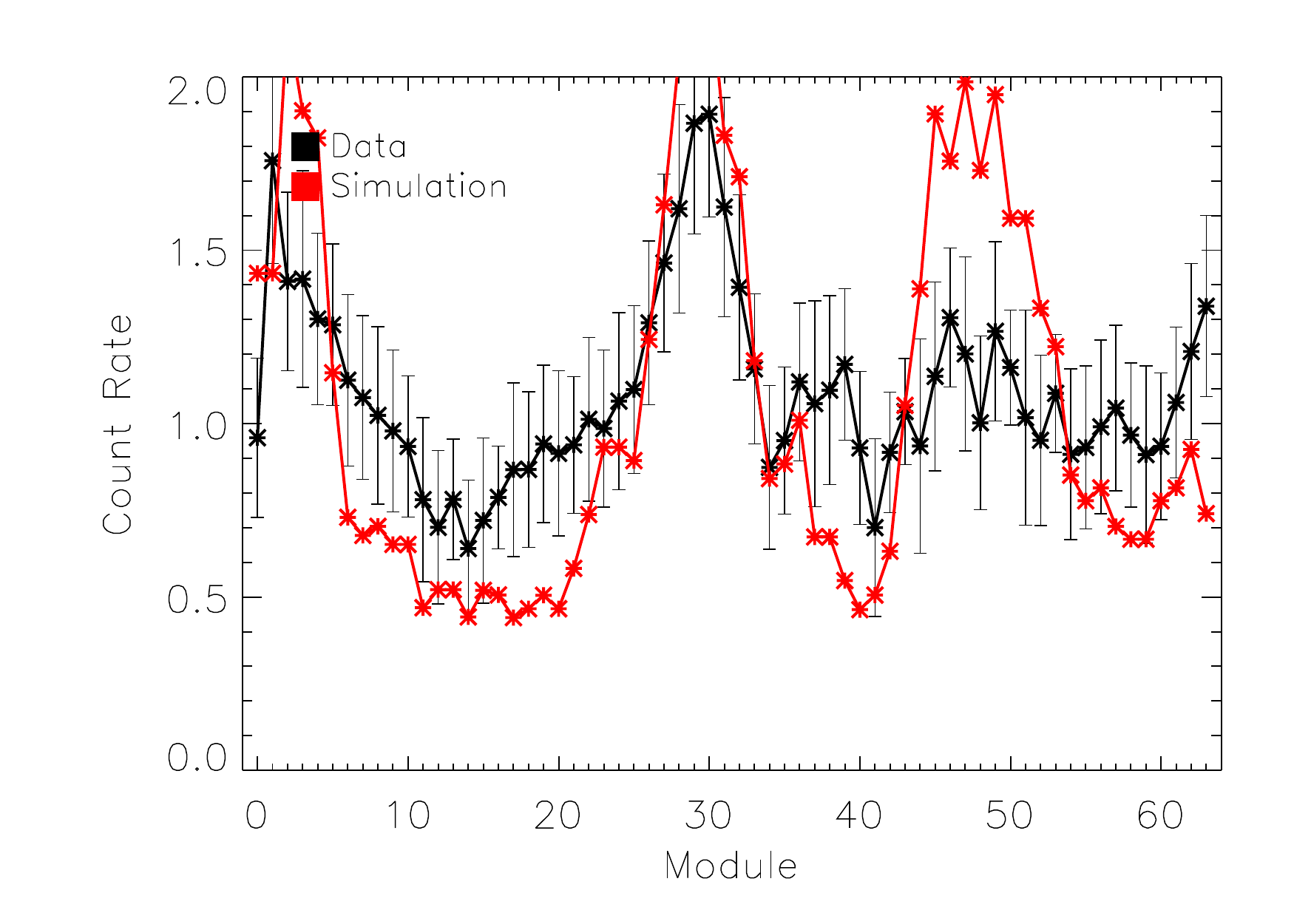}
\caption{GRB 190530A: $\theta = 154.50^\circ$, $\phi = 79.87^\circ$}
\end{subfigure}
\caption{Comparison between observed (black) and simulated (red) detector plane histogram (DPH) of 2-pixel Compton events binned at module level for a sample of  GRBs.
}
\label{fig_comp_dph}
\end{figure*} 

In a pixelated detector like CZTI, where the pixels are defined by the anode pattern, it is likely that any charge deposited close to the pixel boundary is shared between two adjacent pixels. The effect of such charge sharing is expected to increase at higher energies because of larger track lengths of the high energy electrons resulting in larger
radius of the charge cloud. The charge sharing events in which the leaked charge is above the energy threshold of the neighboring pixel, results in double pixel events and can mimic the Compton event. The effect of charge sharing can be seen in Fig. \ref{fig_events}, where
we compare the ratio of single pixel events to 2-pixel Compton events seen in the observed data and Geant4 simulations separately in 100-600 keV for all the twenty GRBs. The ratio obtained from simulation (red `cross') is consistently higher than the observed values (black `triangles'). 
We try to verify that this difference is a result of charge sharing by estimating the area of charge sharing region in each pixel and use that fractional area ($f$) to correct the number of single ($N_{single}-f\times N_{single}$) and 2-pixel Compton events ($N_{Compton}+f\times N_{Compton}$).
This fractional area is estimated from the size of the charge cloud at the electrode which in turn is calculated from the detector depth, charge drift distance, detector substrate bias, and temperature  \citep[CZTI charge sharing calculations are given in detail in][]{chattopadhyay16}.  Final charge cloud radius also depends on the initial cloud radius which is normally approximated as the range of the photoelectron. For that, we calculate the weighted range of photoelectrons in CZT for the known power law spectra of the GRBs.

In the azimuthal angle distributions, charge sharing effect artificially increases the number of edge double pixels compared to corner double pixels, because in case of sufficient charge leaking into the adjacent corner pixel, at least one (or possibly two) adjacent edge pixels will get higher charge, and hence such event will get registered as three or four pixel event and thus will get rejected during the Compton event selection.
Since the azimuthal histograms obtained by Geant4 simulations are used for geometry corrections of the observed azimuthal histograms, it is necessary to correct it for the charge sharing, which is done by normalizing the edge to corner pixel ratio for the Geant4 simulations to that of the observed data. This correction is applied before the geometry correction, and thus preserves the polarization induced modulation within the edge and corner pixels. We used the same method in experimental analysis and found the results to agree well (Vaishnava et al. 2022, under review).

\subsubsection{Improvements and validation of the {\em AstroSat} mass model}\label{compton_dph}

Polarization for off-axis sources with CZTI critically depends on the accuracy of the mass model of {\em AstroSat}. In order to understand the effect of the surrounding material on unpolarized and polarized radiation, we modeled the entire {\em AstroSat} observatory inside Geant4 including all the payloads of
{\em AstroSat}: SSM, UVIT, SXT, LAXPC, CZTI and the satellite bus. In our previous GRB polarimetry paper, \citet{chattopadhyay19}, we discussed the basic concepts in the development of the mass model and its validation using observed data for GRB 160821A. In recent times, we made a number of changes in the mass model geometries based on the feedback from observed data, particularly, in the CZTI housing and the spacecraft. These details along with the complete {\em AstroSat} mass model development in Geant4 can be found in \citet{mate21}.
In \citet{chattopadhyay21_grb}, we performed broadband spectroscopy for a sample of GRBs using CZTI single pixel and 2-pixel Compton events and using response matrices obtained from Geant4 simulations of the mass model interactions. The spectroscopic results closely agree with the Fermi parameters which validates the mass model.   

The validation of the mass model using a large sample of GRBs by comparing the simulated and observed detector plane histograms (DPH) of single pixel events in 70-200 keV is also attempted in \citet{mate21}. It was, however, found that the simulation could not fully replicate all the observations, perhaps due to the fact that the single events contain leaked low energy photons through small sized gaps (which could not be adequately modeled), along with the scattered photons.  
Hence, in order to validate the mass model further particularly in the context of polarization analysis, here we attempt to compare the simulated DPHs with the observed DPHs generated from the same Compton events which are used in polarization analysis. 
The analysis was performed for all the twenty GRBs reported in the present work using Geant4 simulations. 
The simulation is done for a large number of photons (10$^9$) with the same spectral energy distribution observed for the GRBs. In the simulation, we implement the observed LLDs and ULDs of the CZTI pixels including the low gain pixels. When the surrounding material and the spacecraft see an incoming photon, the outgoing energy and direction of the photon depend on its initial energy and momentum direction.  
We generated DPHs for all the twenty GRBs in multiple energy bands and found that the simulated and observed DPHs for all the GRBs are found to agree well in all the energy bands. 
Figure \ref{fig_comp_dph} shows the observed and the simulated module-wise DPH for a sample of representative  GRBs in the 100-600 keV band.

\section{Results}\label{results}
\begin{figure}
\centering
\includegraphics[scale=.6]{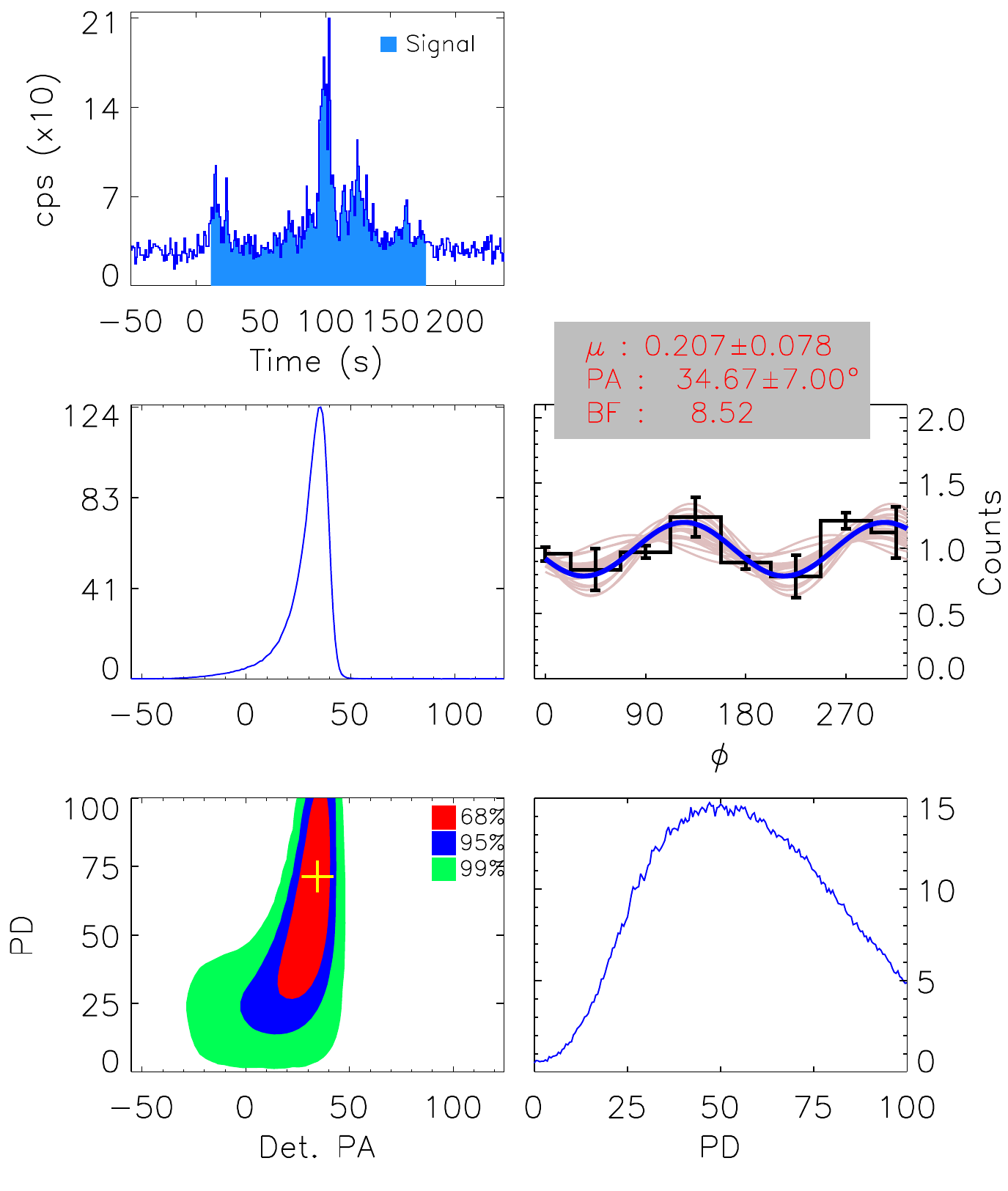}
\caption{Modulation curve for GRB 180103A. Top: 1 second Compton light curve in 100-600 keV. The Compton events for polarization analysis are obtained from the blue shaded region. Middle left: posterior probability distribution for polarization angle from MCMC iterations. Middle right: the modulation curve and the sinusoidal fit shown in solid blue line, along with  100 random MCMC iterations. 
Bottom left: the contour plot for polarization angle and degree for 68, 95,  and 99\% confidence levels. Bottom right: posterior probability distribution for polarization degree from MCMC iterations. Similar figures are given for the remaining 19 GRBs in the online journal.}
\label{fig_mod_180103A}
\end{figure}

For time and energy integrated polarization, the Compton events in the full burst interval in the complete energy range of 100-600 keV are used.  
The background subtracted and geometry corrected azimuthal angle distribution of the valid Compton events are fitted with a sinusoidal function, $A\cos2(\phi - \phi_0+\pi/2)+B$, where A and B determine the modulation amplitude of the signal, $\mu = A/B$ and $\phi_0$ is the polarization angle in the detector plane. 
To estimate the fitting parameters (A, B, $\phi_0$) and the associated errors, we perform MCMC 
simulations for a large number (1 million) of iterations. For each iteration, the posterior probability is estimated based on randomly sampled model parameter values.
Figure \ref{fig_mod_180103A} shows the posterior probability distribution for $\mu$ (bottom right plot) and polarization angle in the detector plane, $\phi_0$ (middle left plot) for one of the bursts, GRB 180103A.
The 1 s light curve and the selected region (based on Bayesian block analysis, see Sec. \ref{sample}) are shown in the top panel of the figure. The 8-bin modulation curve after background subtraction and geometry correction along with the sinusoidal fit (solid blue line) is shown in the middle right plot. We also show a distribution of the sinusoidal fits for 100 random iterations in solid pink lines.      
The 68\%, 95\% and 99\% contours for polarization fraction and angle for GRB 180103A are shown in the bottom left plot of Fig. \ref{fig_mod_180103A} in red, blue and green respectively.

\begin{table*}
\begin{center}
\caption{Measured polarization fractions (PF) and position angles (PA) for the GRBs in (100-600 keV) energy range}
\begin{tabular}{ c c c c c c  }
 \hline
GRB Name  & N$_{compt}$  & Bayes Factor & PF ($\%$)$^a$ & CZTI PA ($^\circ$)$^b$ & sky PA ($^\circ$)  \\
\hline
\hline
 GRB 160325A &  764 & 1.72 & $<45.02$ & --  & -- \\
 GRB 160623A & 1714 & 1.02 & $<56.51 $ & -- & -- \\
 GRB 160703A &  433 & 0.76 & $<62.64 $ & -- & -- \\
 GRB 160802A & 1511 & 0.69 & $<51.89 $ & -- & -- \\
 GRB 160821A & 2851 & 0.87 & $<33.87 $ & -- & -- \\ 
 GRB 170527A & 1638 & 0.79 & $<36.46 $ & -- & -- \\
 GRB 171010A & 3797  & 0.98 & $<30.02 $ & -- & -- \\ 
 GRB 171227A & 1249  & 0.84 & $<55.62 $ & -- & -- \\
 GRB 180103A & 4164 & 8.52 & $71.43 \pm 26.84$ & $34.67 \pm 7.00$ & $122.13$ \\
 GRB 180120A &  705 & 3.95 & $62.37 \pm 29.79$ & $-3.65 \pm 26.00$ & $61.21$ \\
 GRB 180427A &  986 & 9.25 & $60.01 \pm 22.32$ & $16.91 \pm 23.00$ & $47.22$ \\
 GRB 180806A &  555 & 0.86 & $<95.80 $ & -- & -- \\
 GRB 180809B & 3294 & 0.98 & $<24.63 $  & -- & -- \\
 GRB 180914A & 2276 & 1.2 & $<33.55 $ & -- & -- \\
 GRB 180914B & 7765 & 3.52 & $48.48 \pm 19.69$ & $26.99 \pm 19.00$ & $68.41$ \\
 GRB 190530A & 1859 & 3.08 & $46.85 \pm 18.53$ & $43.58 \pm 5.00$ & $154.05$ \\
 GRB 190928A & 4492 & 1.77 & $<33.10$ & -- & -- \\
 GRB 200311A & 1082 & 0.86 & $<45.41 $ & -- & -- \\
 GRB 200412A &  911 & 0.89 & $<53.84 $ & -- & -- \\
 GRB 200806A &  534  & 0.71 & $<54.73 $ & -- & -- \\
\hline
 \end{tabular}
 \label{table_final}
 \end{center}

a: the upper limits are calculated at 2$\sigma$ level and the error bars are at 1$\sigma$ level for the 1 parameter of interest.\\
b: the error bars are at 1$\sigma$ level for the 1 parameter of interest.
\end{table*}
In order to claim that the burst is truly polarized, we estimate Bayes factor for the sinusoidal model (M$_1$, for polarized photons) and a 
constant model (M$_2$, unpolarized photons) as the ratio of marginal likelihoods of M$_1$ to M$_2$ \citep[for more details, see][]{chattopadhyay19}.  
For the GRBs with the Bayes factor greater than 3, we estimate polarization fraction and angle from the fitted parameters.  
For example, for GRB 180103A, the Bayes factor is found to be 8.35, implying that the signal is truly polarized and probability of an unpolarized radiation mimicking such sinusoidal modulation in the azimuthal angle distribution is low \citep[$<$0.1\%, see][for details of chance probability calculations]{chattopadhyay19}. 
Polarization fraction is estimated by normalizing the fitted modulation amplitude, $\mu$ by modulation
factor for 100$\%$ polarized radiation ($\mu_{100}$), which is calculated by simulating the {\em AstroSat} mass model in Geant4.  
Table \ref{table_final} summarizes the results for all the twenty GRBs\footnote{full data set and analysis techniques would be made available based on reasonable requests}. 
The estimated Bayes factors are given in the third column of the table. We see that besides GRB 180103A, four other GRBs (GRB 180120A, GRB 180427A, GRB 180914B, and GRB 190530A) have Bayes factor above 3 for which polarization fractions are estimated (fourth column). The fitted polarization angles in the CZTI and in sky plane (North to East in anti-clockwise direction) are given in the fifth and sixth columns of Table \ref{table_final} respectively for these five GRBs. In the second column, we give the number of selected Compton events in the full burst intervals of the GRBs. 

It is to be noted that the uncertainties quoted in Table \ref{table_final} on polarization fraction and angle 
are obtained at 68\% confidence level for only one parameter of interest, that is by 
looking only at the variation in the azimuthal angle distribution rather than 
the measurement of both polarization fraction and angle simultaneously. The latter is 
resorted to while determining the contours.
For the remaining fifteen GRBs with Bayes factor $<$3, we calculate upper limits in polarization fraction by estimating the polarization detection threshold. Polarization detection threshold is determined 
by limiting the probability of false detection which is 
estimated by simulating unpolarized incident radiation in Geant4 using the {\em AstroSat} mass model for the 
observed Compton and background events for a given GRB. The method of upper limit estimation is described in \citet{chattopadhyay19}.

From Table \ref{table_final}, we see that only 25\% of the sample is polarized according to the convention that the signal is polarized if the Bayes factor is greater than 3. The polarization level ranges between $\sim$50 and 70\%. For a significant fraction of GRBs, the 2$\sigma$ polarization upper limits are below 40\%, signifying that most of the GRBs are unpolarized or polarized at a low level.   
It is to be noted that since the upper limit or the false detection probability is also estimated by looking only at the variation in the azimuthal angle 
distribution for unpolarized radiation, for certain bursts, even though the polarization
contours are unconstrained at 68\% level, we still obtain meaningful upper limits on polarization fraction.

\section{Discussions}\label{discussion}

In this work, we have presented the results of a comprehensive analysis of the GRBs detected by CZTI on board the {\em AstroSat} satellite and have given the hard X-ray polarization results for twenty GRBs. In this section, we compare these results with those obtained by POLAR \citep{Kole20polar_catalog}.
We also  discuss the implications of our polarization results in terms of emission mechanism of the prompt bursts and generic features of GRB prompt emission polarization. 
We summarise the improvements that we have made in the polarization analysis technique and sketch possible future improvements and observation strategies.

\subsection{Comparison with POLAR results}\label{compare}

\citet{Kole20polar_catalog} using POLAR reported low polarization for time integrated emissions of a sample of fourteen GRBs with only two GRBs having polarization fraction above 40\%.   
In \citet{chattopadhyay19}, we reported polarization measurements for a sample of eleven bright GRBs detected in the first year of {\em AstroSat}. However, these measurements were not carried out in the full burst interval of the GRBs, rather the selection of the intervals and the energy range were optimized to obtain the best detection of polarization. Most of the GRBs were found to have high polarization. 
In this work, however, polarization analysis has been done systematically for the full burst duration in 100-600 keV for all the GRBs in the sample. Since both POLAR and CZTI measurements reported here are carried out for the full burst intervals, we expect similar polarization properties in both cases. Two out of fourteen POLAR GRBs were reported to have polarization above 40\%, which translates to 14\% of the sample. In case of CZTI, for five GRBs, polarization fraction and angles were constrained with polarization fraction above $\sim$50\%, implying that around 25\% of the GRBs are polarized. 

\begin{figure}
\centering
\includegraphics[scale=.43]{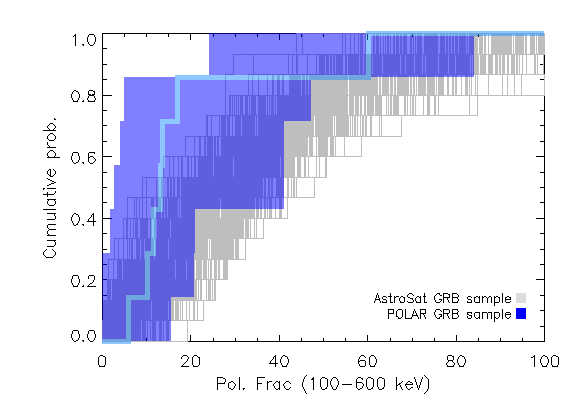}
\caption{Comparison of POLAR and CZTI polarization results. Cumulative distribution of polarization fractions reported by POLAR for seven GRBs with fluence above 10$^{-5}$ erg-cm$^{-2}$ (10-1000 keV) is shown by a solid blue line with 1 $\sigma$ error boundary. The gray lines are obtained from fifteen GRBs detected by CZTI with fluence above 10$^{-5}$ erg-cm$^{-2}$ (10-1000 keV) and at angles less than 60$^\circ$ (see text for more details).}
\label{fig_cum_dist}
\end{figure}

We used the reported polarization fraction values for the POLAR GRBs and the associated 1$\sigma$ errors from Table 2 of \citet{Kole20polar_catalog} and computed the cumulative probability of polarization fraction. This is shown in Fig. \ref{fig_cum_dist} in blue along with the 1$\sigma$ error boundaries. 
A similar distribution of cumulative probability of polarization fraction for the CZTI GRB sample is shown in gray. Each distribution is computed from randomly sampled values drawn from a Poisson distribution around the estimated polarization fraction with error on fraction as the standard deviation for the GRBs with Bayes factor above 3. In case of GRBs with Bayes factor less than 3, we randomly sample a uniform distribution between 0 and the respective polarization upper limits. 
In order to avoid any bias in the sample, we select only the GRBs with fluence above 10$^{-5}$ erg/cm$^2$ in 10-1000 keV band for both POLAR and CZTI. In addition to that, for CZTI, we selected the GRBs that are detected at incident angles $\lesssim$60$^\circ$ as the sample is complete in this angle range (see discussion in Sec. \ref{sample}).  
We see that 
CZTI and POLAR show a similar distribution of polarization fraction such that 50\% of the observed GRBs show polarization fraction $<$~50\% of the peak value.
There is, however, some discrepancy in the detected polarization fraction such that either POLAR polarization fractions are systematically lower by around 25\%  or CZTI values are systematically higher by around 25\%. 
We expect a certain level of systematic error associated with both POLAR and CZTI or any polarimetry instrument in general.
However, CZTI (or POLAR) known (or identified) instrumental systematics alone are unlikely to cause a discrepancy of 25\%.
We note that POLAR samples a wider time range because of its high sensitivity down to 50 keV. CZTI, on the other hand, being highly sensitive in the higher energy range above 100 keV, gives a snapshot picture of the narrow peak in the high energy.
We quantified this by conducting Bayesian block analysis for the common 13 GRBs detected by CZTI and {\em Fermi} in 50-500 keV, the energy range where POLAR is sensitive. 
We found that the estimated burst intervals are around a factor of 2 higher in 50-500 keV than those estimated in 100-600 keV similar to that estimated in CZTI.
GRB prompt emission is highly structured leading to possible changes in the polarization angle within a burst which has been seen for a number of GRBs, e.g., GRB 160821A \citep{Sharma_etal_2019}, GRB 170114A \citep{zhang19}, and GRB 100826A \citep{yonetoku11}.
Thus, we surmise that the sampling of larger time intervals 
by POLAR may have led to the dilution of the polarization fraction across the burst duration, in comparison to the narrower time intervals sampled by CZTI.   

We have also attempted polarization measurement for GRB 161218B which was detected by both CZTI and POLAR. 
Because the GRB is faint, POLAR could not constrain the polarization parameters. POLAR reported a polarization degree, $\rm PD = 13^{+28}_{-13}\%$ and sky polarization angle, $\rm PA = {68^{+36}_{-54}}^{\circ}$ \citep{Kole20polar_catalog}.
The incident direction of GRB 161218B in CZTI was $\theta=121.63^{\circ}, \phi=117.71^{\circ}$.
For the polarization and spectral analysis of GRB 161218B with CZTI, we chose the same region of the burst selected by POLAR for polarization analysis with burst interval around 25.1~s.
The number of Compton counts in the burst duration was found to be 418 (below our selection criteria for inclusion in the 5-year sample).  
We performed the standard polarization analysis for this GRB and found the Bayes factor to be low, consistent with zero polarization (in agreement with POLAR result). 
 
\subsection{Correlation between polarization and spectral parameters} \label{correlation}
\begin{figure*}
\centering
\begin{subfigure}[b]{0.4\textwidth}
\includegraphics[width=\textwidth]{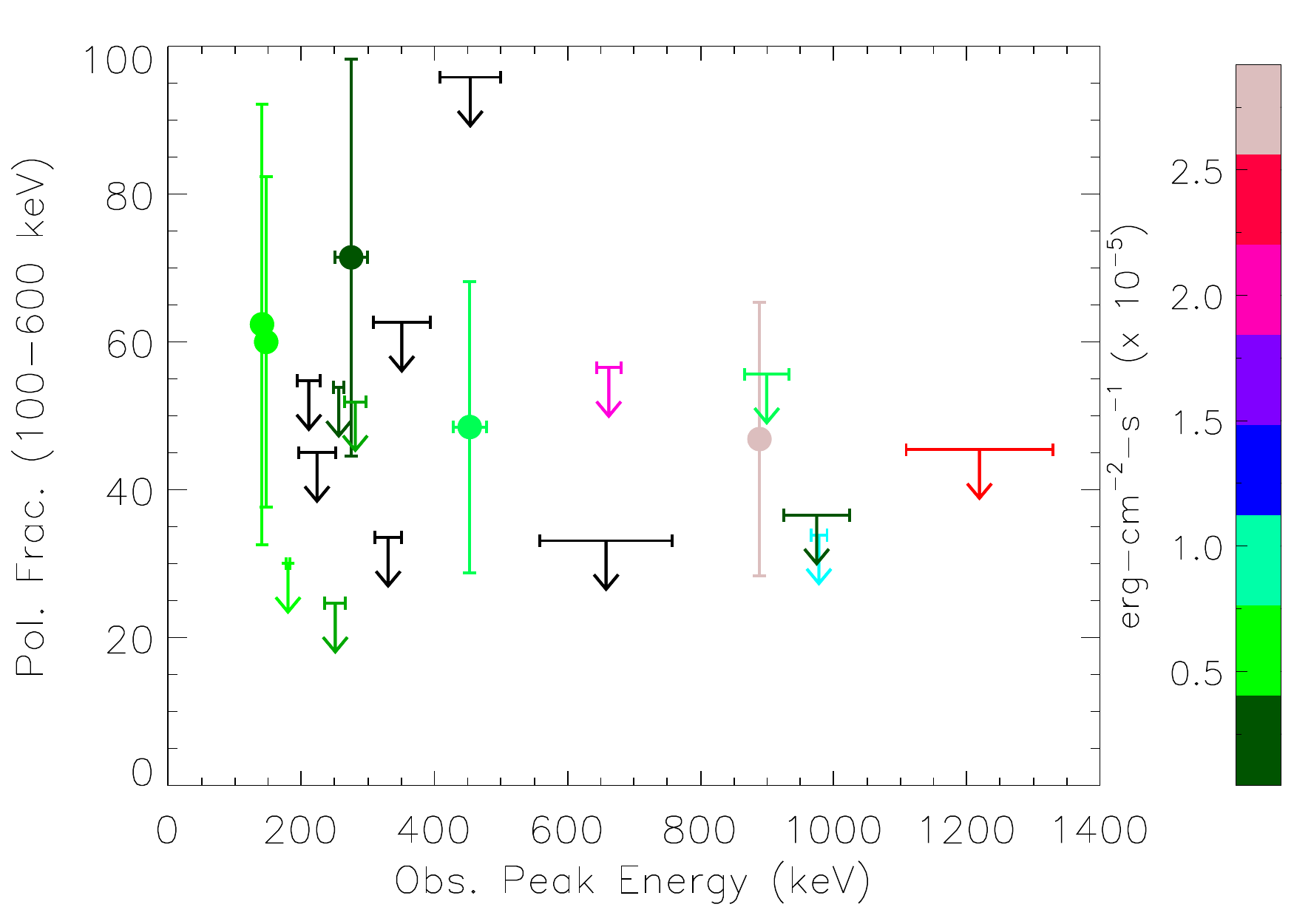}
\caption{}
\end{subfigure}\\
\begin{subfigure}[b]{0.4\textwidth}
\includegraphics[width=\textwidth]{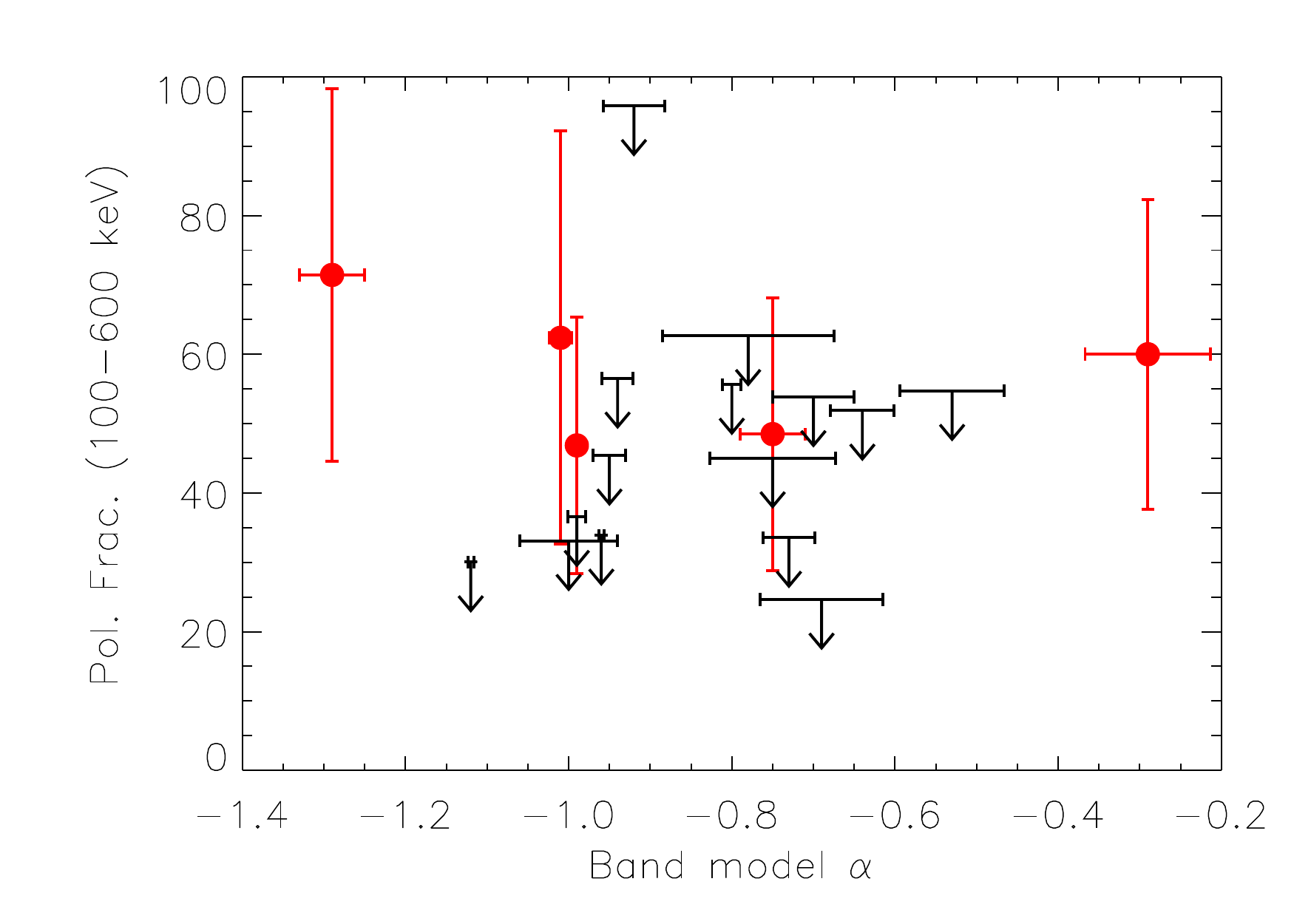}
\caption{}
\end{subfigure}
\begin{subfigure}[b]{0.4\textwidth}
\includegraphics[width=\textwidth]{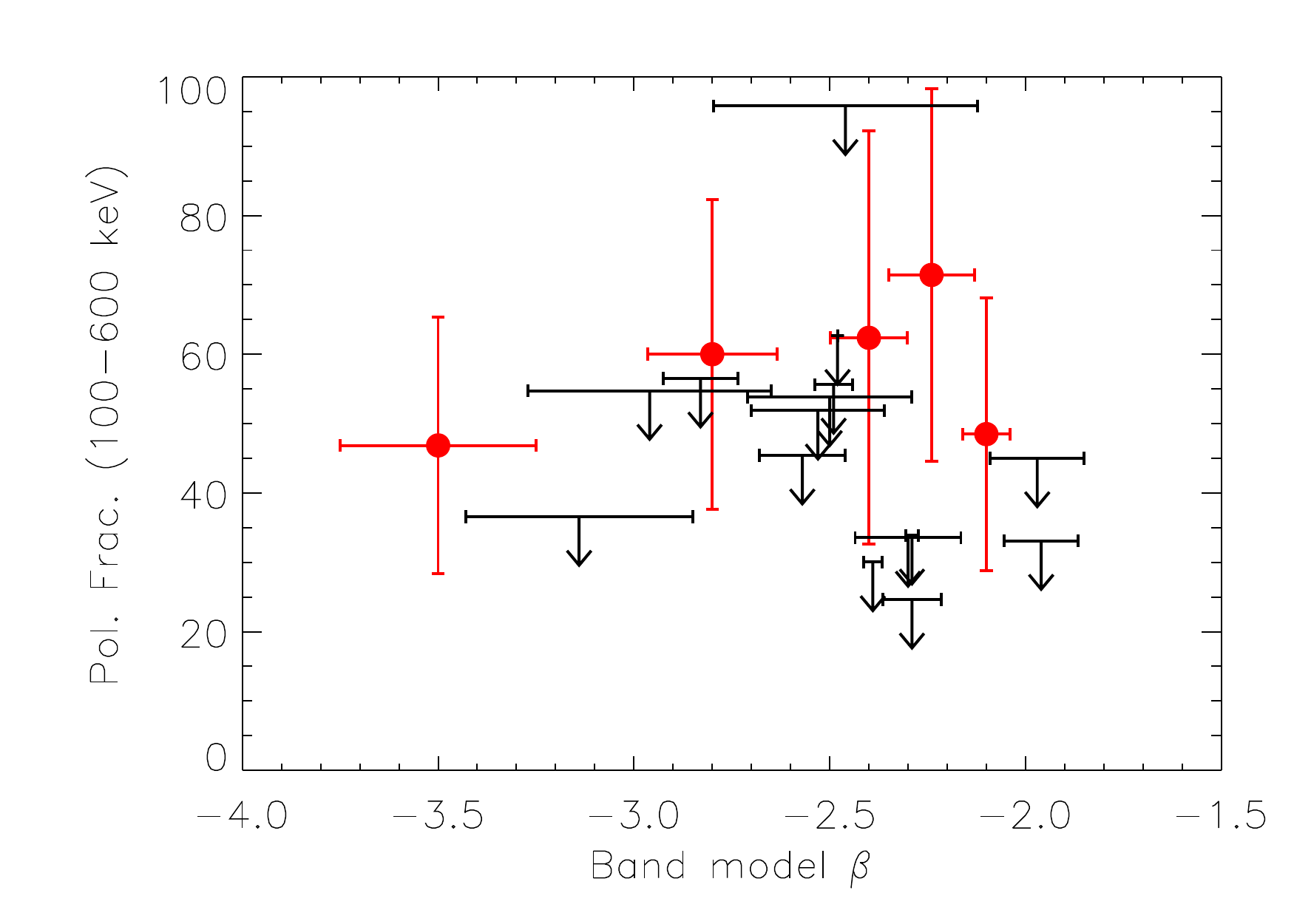}
\caption{}
\end{subfigure}
\caption{(a) Polarization fraction of the GRBs as a function of peak energies, $E_{\rm peak}$. The 10-1000 keV fluences in $erg-cm^{-2}-s^{-1}$ for the GRBs are color coded.
(b),(c) Polarization fractions of the GRBs are shown with $\alpha$ (b)  and  $\beta$ (c)  obtained from Band function fits to the time integrated burst spectra. The red circles stand for GRBs with Bayes factor above 3 whereas the black down arrows represent the GRBs with Bayes factor less than 3.
}
\label{fig_epeak_PF}
\end{figure*}

Different radiation models for the prompt emission of GRBs predict different ranges of possible polarization fractions depending on the configuration of magnetic fields at the emitting site, the viewing geometry, and the jet structure \citep[see Fig. 1 in][]{Gill_etal_2021_review}. It is not possible to spatially resolve the emitting region and thereby the observed polarization measurement is always an average of the radiation coming from the visible emitting region. 

Synchrotron emission produced from a distribution of electrons is intrinsically linearly polarized. However, depending on the configuration of magnetic field within the 1/$\gamma$ view cone ($\gamma$ being the bulk Lorentz factor of the jet), the average observed polarization can vary.  Viewed within the jet opening angle $\theta_j$, a random orientation of the magnetic field would always produce low to null polarization and so will an azimuthally ordered magnetic field if the view axis coincides with the jet axis \citep{toma08}.  The latter magnetic geometry may however produce significant polarization, between 16 to 70\%, for spectral slope $\alpha$ in the range $-0.67$ to $-1.5$ and off-axis view geometry with $y_j \equiv (\gamma \theta_j)^2$ equal to 1 or higher for a top hat jet \citep{granot03}.  On the other hand, Photospheric emission, primarily resulting from Compton scattering, would exhibit little polarization within $\theta_j$ \citep{toma08}.  View angles outside the jet cone can lead to high observed polarization due to emission zone asymmetry \citep{granot03,toma08,lundman14}, but in such cases the observed flux would be much lower and may preclude the measurement of polarization.  
In case of subphotospheric dissipation too the observed emission can be polarized, but only at energies much below the spectral peak \citep{Lundman_etal_2018}. 

The population properties of GRB polarization can also act as a possible discriminator between emission models. \cite{toma08} demonstrate that for synchrotron emission in ordered magnetic fields, a negative correlation between polarization fraction (PF) and spectral peak energy ($E_{\rm peak}$) may be expected, while this is not so in other emission models.
With this in mind, we have plotted the measured PF and upper limits (100-600 keV) of our GRB sample against the observed $E_{\rm peak}$ in Fig. \ref{fig_epeak_PF}, color coded according to the estimated 10-1000 keV burst flux. 
Well constrained PF values are obtained for five cases, with PF lying in the range $\sim$28--97\% as per the confidence interval of one parameter of interest. These bursts have fluence values among the highest observed, which suggests that they are likely viewed within the jet opening angle. If so, then the strong observed polarization would prefer synchrotron emission in ordered magnetic fields as the main radiation mechanism.
However, to confirm this assessment via a PF-$E_{\rm peak}$ correlation, one would require a larger sample of GRBs than we currently have.

In Fig. \ref{fig_epeak_PF}b and c, we also plot PF versus $\alpha$ and $\beta$ of the time integrated burst spectra of our GRBs. No significant trend is observed between PF and the spectral slopes. 
Interestingly, we note that one GRB (180427A) exhibits a high polarization of 37-83\% while its time integrated spectrum has a hard $\alpha \sim -0.3$ and steep $\beta \sim -2.8$. The spectrum is suggestive of photospheric emission, in which case a high polarization can be observed only when viewed outside the jet opening angle \citep{waxman03,chand18a}. This case emphasizes that high PF alone cannot provide a full diagnostic of the emission mechanism, rather a spectro-polarimetric study is required to get a better handle. 

As discussed earlier, the PF could not be constrained for a majority of the bursts in our sample, indicating a low to null average polarization over the burst duration in these cases. This finding is consistent with that reported by POLAR for their sample of the study. 
The observed PF averaged over the burst can remain low to null due to either the radiation being intrinsically unpolarized or the angle of polarization varying temporally or with energy across the burst emission. 

\subsection{Improvements in the polarization analysis techniques} \label{improvement}

Compared to the earlier work on GRB polarization using CZTI, we have made the following significant modifications in the analysis technique -- a) inclusion of low gain pixels and thus increasing the effective area and the energy range, b) correcting for 2-pixel events generated by  charge sharing and c) noise reduction. 
We find that the Compton noise reduction has a marginal impact on the results for most of the GRBs. In a few  cases where the number of noisy events is large, 
the excess events arising from the pairs of noisy pixels might enhance counts in certain azimuthal bins and thereby can give artificially higher amplitude of modulation. 

The correction for charge sharing, on the other hand, is extremely important in removing systematic effects in the modulation curves. As mentioned earlier, the ratio of edge to corner pixel bins is lower in simulations if we do not consider charge sharing. In that case, even after geometry correction for a completely unpolarized radiation, the modulation curve would still not be flat with the edge bins and corner bins separated by a statistically significant amount. A sinusoidal fit to this modulation curve would yield a lower amplitude or polarization fraction but in general with a high detection significance as the values in alternate bins are often outside the error bars of each other. We find that  without charge sharing correction, the estimated polarization fractions are lower but with slightly better detection significance. It is important to account for charge sharing effect in simulation to remove these systematics from the modulation curves. 

The inclusion of the low  gain pixels increases the number of Compton events by 30-40\% and, more importantly, extends the energy range to 600 keV. We find that the background is higher at higher energies and hence the resultant improvement in the detection significance is marginal. The extension of the energy range, however, makes the results robust and will help us in possible future energy resolved spectro-polarimetry.

We have re-analyzed all the published  polarization results based on CZTI data using the new method and find that the results are consistent with the published ones, albeit with reduced significance.
The upper limits as derived with the new method are consistent with those published earlier.  However, for GRB 160802A \citep{chattopadhyay19} and GRB 160821A \citep{Sharma_etal_2019}, we see the detection significance in polarization fraction has been reduced in the new method. The apparent lower errors in the older method resulted from the absence of charge sharing correction. 
For GRB 160821A, even with the new method, we find the $\sim$90$^\circ$ change in the polarization angle across the first and second parts and second and third parts of the prompt emission, albeit with reduced  significance.  

\section{Conclusions} \label{Conclusions}

The polarization property of GRB prompt emission is still a less explored field among the astronomical observations. It is highly challenging due to the transient nature of GRBs resulting in poor photon statistics and also due to the involvement of non-trivial instrument systematics. The comprehensive analysis presented here for the  five year data from CZTI, in conjunction with the polarization results presented for a  sample of GRBs using POLAR \citep{zhang19,Kole20polar_catalog}, demonstrate that a large fraction of GRBs show low polarization. There exists, however, a significant fraction of GRBs ($\sim$14-25\%) showing high hard X-ray polarization ($\ge$40\%). In these samples the number of detections so far is too low to attempt correlation studies with other observed properties of the prompt emission. On the other hand, the high polarization fraction in a few GRBs, with a hint of variability in the polarization fraction and polarization angle, offers an opportunity to study these variations and associate them with the other observables. We plan to undertake a systematic exploration of time resolved spectro-polarimetry of the GRBs with high polarization. CZTI is in its seventh year of operation and it continues to provide valuable information on GRB polarization.  We plan to conduct detailed prompt and after-glow emission measurements of bright GRBs with a view to associate various observed properties with polarization variations.
In future, dedicated polarimetric instruments such as LEAP \citep{mcconnell21_leap}, POLAR-2 \citep{kole19}, Daksha\footnote{\url{https://www.star-iitb.in/research/daksha}} (a multi-messenger astronomical mission proposed to the Indian Space Research Organization) etc, will enable more such spectro-polarimetric measurements.

\begin{acknowledgments}
This publication uses data from the {\em AstroSat} mission of the
Indian Space Research Organization (ISRO), archived at the
Indian Space Science Data Centre (ISSDC). CZT-Imager is built
by a consortium of Institutes across India including Tata Institute
of Fundamental Research, Mumbai, Vikram Sarabhai Space
Centre, Thiruvananthapuram, ISRO Satellite Centre, Bengaluru,
Inter University Centre for Astronomy and Astrophysics, Pune,
Physical Research Laboratory, Ahmedabad, Space Application
Centre, Ahmedabad: contributions from the vast technical team
from all these institutes are gratefully acknowledged. DF acknowledges support from RSF grant 21-12-00250.
This research has also made
use of data obtained through the High Energy Astrophysics
Science Archive Research Center Online Service, provided by the NASA/Goddard Space Flight Center.
\end{acknowledgments}





\end{document}